\documentclass[twocolumn,showpacs,preprintnumbers,amsmath,amssymb]{revtex4}

\usepackage{graphics}

\usepackage{epsfig}

\begin{document}


\title{Melting curve of $^4$He: no sign of the supersolid
transition down to 10\,mK}


\author{I.\,A.\,Todoshchenko}
\author{H.\,Alles}
\author{J.\,Bueno}
\altaffiliation{Kamerlingh Onnes Laboratory, Leiden University,
P.\,O.\,Box 9504, Leiden, 2300 RA Netherlands}
\author{H.\,J.\,Junes}
\author{A.\,Ya.\,Parshin}
\altaffiliation{P.\,L.\,Kapitza Institute, Kosygina 2, Moscow
119334, Russia}
\author{V.\,Tsepelin}
\altaffiliation{Department of Physics, Lancaster University,
Lancaster, LA1 4YB}
\affiliation{Low Temperature Laboratory, Helsinki University of
Technology, P.\,O.\,Box 2200, FIN-02015~HUT, Finland}

\email[]{todo@boojum.hut.fi}

\date{\today}

\begin{abstract}
We have measured the melting curve of $^4$He in the temperature
range from 10 to 400\,mK with the accuracy of about 0.5\,$\mu$bar.
Crystals of different quality show the expected $T^4$-dependence
in the range from 80 to 400\,mK without any sign of the
supersolid transition, and the coefficient is in excellent
agreement with available data on the sound velocity in liquid
$^4$He and on the Debye temperature of solid $^4$He. Below 80\,mK 
we have observed a small deviation from $T^4$-dependence which
however cannot be attributed to the supersolid transition because
instead of decrease the entropy of the solid rather remains 
constant, about $2.5\times10^{-6}$\,$R$. 
\end{abstract}




\maketitle

In 1969 Andreev and Lifshitz have proposed that owing to the 
large zero-point motion of the atoms, helium crystals may 
contain a finite concentration of vacancies even at absolute 
zero temperature \cite{AL}. These vacancies are
delocalized and do not violate the crystalline symmetry. At 
sufficiently low temperature the vacancies may Bose-condense, 
in which case the crystal phase would manifest quite unusual 
properties, such as a non-dissipative mass current.

This possible ``supersolid'' state was a subject of great interest
in the 1970s and 1980s, when several experimental groups tried to
detect the superflow of vacancies by various methods (see
\cite{Meisel} for a review). While all these attempts were
unsuccessful, they did put an upper limit for the possible
superfluid fraction at $5\times 10^{-6}$, and for the value of
the critical velocity at $<5\times 10^{-8}$\,cm$/$s down to 25\,mK
\cite{Meisel}. The only exceptions were the ultrasonic experiments
with ultrapure (1.5\,ppb of $^3$He) solid $^4$He by Lengua and
Goodkind \cite{Lengua}, who observed an increase of the sound
attenuation at low temperatures. They attributed this phenomenon
to the interaction between phonons and vacancies, and obtained a
superfluid fraction $\rho_s/\rho\sim10^{-3}$ and a condensation
temperature of the order of 0.1\,K.

Interest in the problem of supersolid was renewed with the
observation by Kim and Chan \cite{Kim} in 2004 of the reduction
in the rotational inertia of a cell containing solid $^4$He below
about 0.2\,K. The authors named this phenomenon
``nonclassical rotational inertia" and interpreted it in terms of
superfluidity of the solid, possibly caused by Bose-Einstein 
condensation of vacancies \cite{Nature}, estimating a superfluid fraction $\rho_s/\rho$ to be from 0.5 to 1.5\,\%, depending on the 
density of the solid and purity of the $^4$He. Their observations 
were recently confirmed by Rittner and Reppy \cite{Rittner}, who 
also pointed out that this effect could only be observed in very 
non-uniform samples grown at relatively high temperatures and then 
rapidly cooled. If such samples are annealed by thermal cycling, 
the effect disappears. However, recent experiments by Kim and Chan
\cite{Kim06} did not confirm the annealing effect.

Most of the previous searches for the supersolid have been via
attempts to detect unusual features in the dynamic behavior of the
solid sample, yielding various interpretations
\cite{Dash,Balatsky,Day}. Our motivation was to check for the
peculiarities of the equilibrium thermodynamical properties of
solid $^4$He, its entropy in particular, which should manifest an
anomaly at the transition (if it really is a phase transition in
the bulk solid). 

The thermodynamical properties of solid $^4$He at relatively high 
temperatures were intensively investigated during the 1960s and 
1970s, and have been thought to be well understood for more than 
30 years. The {\it hcp} solid $^4$He is well described by the ``classical'' concept of phonons, with an effective Debye 
temperature $\Theta\approx26$\,K. The temperature dependence of 
$\Theta$ becomes negligible below 0.5\,K \cite{Gardner}. This 
implies that below 0.5\,K the heat capacity of solid $^4$He varies 
as $T^3$, $C_S=(12/5)\pi^4R(T/\Theta)^3$.
The slope of the melting curve is proportional to 
the difference of the entropies of the liquid and solid,
$dp/dT|_{\mathrm{MC}}=(S_L-S_S)/(v_L-v_S)$, where $v_L$ and $v_S$
are molar volumes of liquid and solid $^4$He. The heat capacity 
of superfluid $^4$He below $\sim0.5$\,K is dominated by phonons,
$C_L=2\pi^2RT^3/(15\hbar^3nu^3)$, where $n$ is the density
and $u=366$\,m$/$s is the sound velocity in liquid $^4$He at 
the melting pressure \cite{Abraham}. Thus the melting pressure 
should vary basically as $T^4$, but should deviate from this 
dependence near any possible supersolid transition. Corrections 
due to rotons, thermal expansion etc.~become negligible below 
0.45\,K \cite{Gardner,Grilly,Greywall79}.

Assuming the superfluid fraction to be about 1\,\% at low 
temperatures, which would correspond to the concentration of 
vacancies of 1\,\% at the condensation temperature, 
one would expect the excess entropy in
the solid of the order of $0.01R$ above the transition
temperature, with a rapid fall in this excess entropy below the 
transition\cite{LL}. For more detailed calculations see 
\cite{Balatsky}.
Such a big excess entropy has never been observed in helium 
crystals. To rule out this discrepancy, Anderson 
{\it et al.}~\cite{Anderson} recently proposed a new model of 
the solid state of $^4$He, where the vacancies and interstitials 
are assumed to be incorporated in a highly-correlated quantum 
state of the crystal, and the only modes giving large contribution 
to thermodynamics are phonons. In this model there is no direct 
connection between the superfluid density and the excess entropy 
above the transition, which behaves as $T^7$ at temperatures up 
to 1\,K and thus may be much smaller than in the case of 
weakly interacting Bose gas. In the view of such a possibility, it 
seems very important to look for any possible anomaly in the 
entropy of solid $^4$He below 0.2\,K with high accuracy.

The heat capacity measurements of solid $^4$He have been carried 
out down to 140\,mK by Castles and Adams \cite{Castles}, who 
have observed a deviation from the $T^3$-behavior already at 
0.4\,K. The melting pressure of $^4$He has been measured
indirectly by Hanson {\it et al.}~\cite{Hanson} who found it to 
vary as $T^4$ down to 100\,mK. Two groups have measured the 
melting pressure of $^4$He at ultra low temperatures. 
Van de Haar {\it et al.}~\cite{Leiden} have found that below 
100\,mK the melting curve is flat with the accuracy of 
20\,$\mu$bar. Measurements with better accuracy by Ruutu 
{\it et al.}~\cite{Ruutu} have shown that the variation of the melting pressure from 100 to 2\,mK is much larger than the 
expected 3.5\,$\mu$bar due to phonons. Unfortunately, Ruutu 
{\it et al.} had very poor thermometry above $\sim10$\,mK.
These results demonstrate that accurate measurements of the 
melting curve of $^4$He in the whole temperature range which 
covers the region of the possible supersolid transition and 
continues down to the lowest temperatures have
been urgently needed.

In this Letter we present our direct high-precision measurements
of the melting pressure of $^4$He in the range from 10 to
400\,mK. With the accuracy of about 0.5\,$\mu$bar we do not see
any deviation from the expected $T^4$-behavior from 80\,mK up to
400\,mK. The variation of the melting pressure obtained,
$-34.2\pm$0.2\,mbar$/$K$^4$, is in good correspondence with the
value of the heat capacity of the solid measured at higher
temperatures, and with the sound velocity in the liquid. We have 
also observed an anomaly below 80\,mK, where the $T^4$-dependence 
changes to much weaker, almost linear dependence.

Our capacitive pressure gauge, of a standard Straty-Adams design \cite{Straty}, is made of beryllium bronze and has the 
sensitivity $dC/dp=44$\,pF$/$bar at the melting pressure 
(25.31\,bar) yielding the accuracy of about 0.5\,$\mu$bar. The 
time and temperature stability measured at zero pressure is of 
the order of $10^{-6}$\,pF. The $^4$He sample, supplied by Oy 
Woikoski AB, Finland, contained less than 0.1\,ppm of $^3$He 
impurities. Temperature was measured by a $^3$He melting curve
thermometer thermally anchored to the sample cell. The
conversion of $^3$He melting pressure to temperature was made
according to the Provisional Low Temperature Scale,
PLTS-2000 \cite{PLTS2000}.

\begin{figure}[t]
\centering
\includegraphics[width=1.0\linewidth]{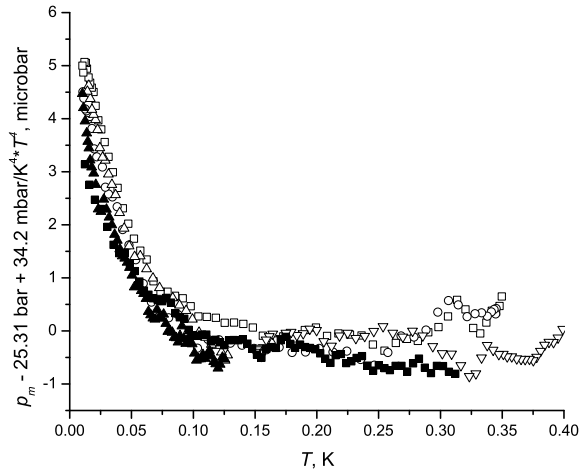}
\caption{\label{fig:t4} Melting pressure of $^4$He 
after substracting the $T^4$-term found from the fit of the 
data at $T>100$\,mK. Data have been taken with the crystals
grown at 1.4\,K and 1.1\,K. Open symbols correspond to cooldowns 
and filled ones to warmups. Different symbols correspond to 
different runs. In the vicinity of the $^3$He melting curve 
minimum at 0.32\,K the scatter of experimental points is 
somewhat larger because of the smaller sensitivity of the 
melting curve thermometer. Note that the substracted 
$T^4$-term varies by about 1\,mbar in this temperature range.}
\end{figure}

The first sample crystal was nucleated and grown at rather
high temperature, 1.4\,K, and then rapidly cooled below 0.6\,K.
Crystals grown in this way are known to contain dislocations,
which facilitates crystal growth at low temperatures, which can
take place without significant overpressures \cite{Ruutu}.
Indeed, our pressure data taken during warming, when the
crystal is growing, do not show any systematic excess over that
taken on cooling (melting). Instead, we have observed a small
hysteresis with opposite sign, which becomes smaller at slower
cooling (warming) rates, apparently due to the relatively high
heat capacity of $^3$He in the melting curve thermometer
(see Fig.\,\ref{fig:t4}).

On the other hand, the melting pressure at 1.4\,K is by about
1\,bar higher than at low temperatures. This pressure
difference leads to large non-uniform stresses in the sample
crystal after cooling down, which may affect the measured
melting pressure \cite{KPS,review}. To check the possible
role of these effects in our measurements, another sample was
grown at $T=1.1$\,K, at a pressure of only $\sim0.1$\,bar 
higher than the melting pressure at low temperatures. Despite 
the fact that such a crystal should be of much better quality, 
no systematic difference was found between the data taken with
this crystal and that of the crystal grown at 1.4\,K. This
demonstrates that the measured melting curve does not depend 
significantly on the concentration of defects in the sample.

Indeed, the contribution of defects like dislocations to the 
melting pressure can be shown to be negligible. At low 
temperatures the entropy of dislocations is due to phonon-like
oscillations of dislocations, 
and can be estimated in the way similar to the calculation 
for usual phonons in bulk solid, 
$S_d\sim\sigma v_Sk_B(T/\Theta_d)/d$. Here $\sigma$ is the 
density of dislocations, $v_S$ is the molar volume, $d$ is the 
lattice constant, and the Debye temperature of dislocations, $\Theta_d$, can be taken the same as for bulk solid.
The concentration of dislocations can be obtained from the
observed threshold for the crystal growth. According to
Ruutu {\it et al.} \cite{Ruutu}, thresholds of the order of
1\,$\mu$bar correspond to the density of dislocations
of about 100\,cm$^{-2}$.
The corresponding additional term to the melting pressure
$\Delta p_{\mathrm{MC},d}\approx10 \sigma k_B(T^2/\Theta_d)/d$
would give a contribution of the order of
$10^{-7}\,\mu$bar$/$K$^2\,T^2$.

The measured melting pressure could also be affected by $^3$He 
impurities. According to the $^3$He-$^4$He phase diagram 
calculated by Edwards and Balibar \cite{Balibar}, the 
concentration of $^3$He in solid $^4$He, $c_{3,S}$, at a 
constant concentration in liquid, $c_{3,L}$, quickly falls 
down below 0.6\,K,
\begin{equation}
c_{3,S}/c_{3,L}\approx exp(-\Delta/T),
~~~\Delta\approx1\,{\rm K}.
\end{equation}
This means that at low temperatures almost all $^3$He atoms
are in liquid, which has much larger volume than the crystal,
and thus the concentration of $^3$He in liquid is the same
as in the whole helium sample, $c_{3,L}<10^{-7}$,
and $c_{3,S}\ll c_{3,L}$. Corresponding contribution to
the slope of the melting curve is positive and less than
10\,$\mu$bar$/$K. We should note, however, that the Eq.\,(1) 
has not been proven experimentally at low temperatures.

\begin{figure}[t]
\centering
\includegraphics[width=1.0\linewidth]{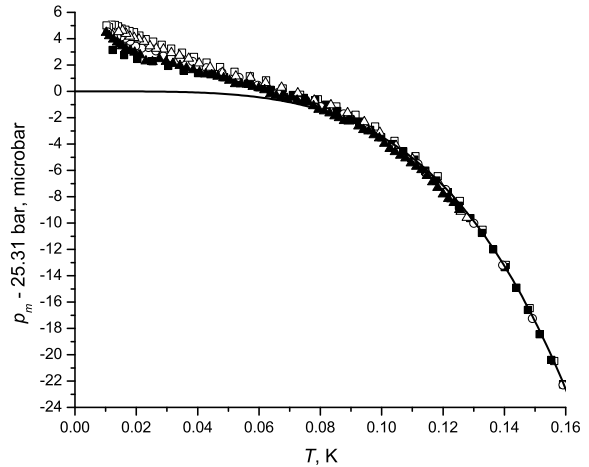}
\caption{\label{fig:lowt} Melting pressure of $^4$He at low
temperatures measured with the crystal grown at 1.1\,K. Open
symbols correspond to cooldown and filled ones to warmup. 
Different symbols correspond to different runs. Solid line
is the $T^4$-fit to the data at $T>100$\,mK extended
down to 0\,K.}
\end{figure}

Our results are shown in Figs.\,\ref{fig:t4} and 
\ref{fig:lowt}. As one can see, the melting pressure measured 
during different runs is reproduced within about 1\,$\mu$bar. 
In the temperature range from 80 to 400\,mK the measured
melting pressure of $^4$He shows the expected $T^4$-behavior with 
the coefficient, $-34.2\pm$0.2\,mbar$/$K$^4$, which is in very 
good agreement with the earlier measurements of melting pressure 
and heat capacity of solid $^4$He at higher temperatures by 
different experimental groups. Indirect measurements of the 
melting pressure by Hanson {\it et al.}~\cite{Hanson} give the variation of the melting pressure -36\,mbar$/$K$^4$, and
from the heat capacity measurements of solid $^4$He by Gardner 
{\it et al.}~\cite{Gardner} the coefficient 
-34.7\,mbar$/$K$^4$ can be found.

Within the accuracy of about 0.5\,$\mu$bar we do not see any manifestation of the transition which Kim and Chan \cite{Kim}, 
and Rittner and Reppy \cite{Rittner} have observed with 
torsional oscillations technique. Kim and Chan have stated that 
their data measured at 26\,bar are consistent with a supersolid 
fraction of about 1\,\% \cite{Kim} and the supersolid transition 
temperature about 200\,mK, but we do not see any anomaly in the 
entropy above 80\,mK with an accuracy of about 
$3\,\times 10^{-7}\,R$ which has to be compared with the expected 
drop of entropy of the order of $0.01\,R$ below the transition 
temperature.

However, below 80\,mK a deviation from $T^4$-dependence
is seen, such that the derivative $-dp/dT$, which is proportional 
to the entropy difference between the solid and liquid, stops decreasing as $T^3$ and saturates at a roughly constant value down 
to our lowest temperature of 10\,mK (see Fig.\,\ref{fig:lowt}). 
Note that the entropy of the liquid is about 5 times less than  
the phonon entropy of the solid, and its contribution
to the melting pressure is small, less than 0.5\,$\mu$bar at
80 mK. Thus, we have an excess entropy in the solid, which
decreases with increasing temperature and disappears around
80 mK. Such rather unusual behavior of entropy (additional
specific heat is negative!) is difficult to understand,
but it certainly cannot be attributed to a supersolid
transition because in this case the additional entropy
would {\it increase} when temperature increases. Also, as it was pointed out in previous paragraph, the amplitude of this anomaly 
is more than four orders of magnitude smaller than expected for 
the Bose-condensation of weakly interacting vacancies.

In the following we discuss possible sources of errors in our 
measurements.
The measured pressure could be influenced by dynamic and
capillary effects and by the hydrostatic pressure change. 
If the crystal quality
is so good that there is no even single dislocation crossing a
facet, then such a facet is practically immobile, and the
corresponding overpressure for its growth may reach
10\,...\,100\,$\mu$bar \cite{Ruutu}. Indeed, when we created
crystals very carefully at low temperatures, 0.6\,K and below,
the measured $^4$He pressure traces typically had significant
temperature hysteresis, up to 10\,...\,20\,$\mu$bar. Measurements 
with the crystal grown at 1.4\,K were much better in this sense, 
with hysteresis of the order of 1\,$\mu$bar and with a tiny drift 
to smaller pressures with time, of the order of
$-2\times 10^{-6}$\,$\mu$bar$/$s. In turn, the crystal grown
at 1.1\,K with better quality, but still with finite amount of
dislocations, showed even better reproducibility and no
measurable time drift. The time drift of the pressure measured
with 1.4\,K crystal we thus interpret as the sign of structural 
disorder in the crystal which slowly becomes smaller. The 
relaxation time of such a process was found to be very long,
of the order of a month. On the other hand, all measured sample 
crystals of different quality (altogether seven) have shown the 
anomaly below 80\,mK.

Capillary contribution to the total equilibrium pressure on the
crystal, being proportional to the surface stiffness $\gamma$,
could reach as much as 10\,$\mu$bar. For a pure $^4$He,
temperature dependence of $\gamma$ is very weak at low
temperatures \cite{AP}, even when taking into account possible
new roughening transitions \cite{review}, and does not contribute
to the slope of the measured melting curve. Due to the adsorption
of $^3$He impurities at low temperatures, $\gamma$ may decrease
by about 20\% \cite{Rolley}, which would produce a decrease
of the measured pressure by a few $\mu$bar at most with
decreasing temperature.

Finally, the change of the hydrostatic pressure due to the
temperature variation is negligible in our experimental
conditions. Indeed, both 1.4\,K and 1.1\,K crystals were grown
up to $\sim$200\,mm$^3$ in volume which means that their lateral
sizes were much larger than the capillary length,
$\lambda\approx1$\,mm. The dependence of the equilibrium height
of such a big ``crystal on a table" on its volume is
exponentially weak, the crystal grows from its sides. The
hydrostatic pressure difference between the crystal and the
pressure gauge thus changes very little due to growth or
melting of the crystal in the course of warming or cooling.
At low temperatures, where the melting curve is almost flat,
this effect is negligible. Moreover, the sign of the effect is opposite to what we observe: when cooling, the crystal melts, 
and the pressure of the liquid in the pressure gauge decreases, 
while we have observed the excess of the melting pressure over 
the expected $T^4$-dependence.

To summarize, the measured melting pressure of $^4$He obeys
the expected $T^4$-law at temperatures above 80\,mK with an
accuracy of about 0.5\,$\mu$bar which is in good agreement 
with previous measurements done at higher temperatures
\cite{Grilly,Hanson,Gardner}. No sign of the supersolid 
transition have been observed. Below about 80\,mK the
$T^4$-dependence changes to a roughly linear
one, which would correspond to about $2.5\times 10^{-6}\,R$
constant entropy in solid $^4$He. However, this anomaly
cannot be attributed to the supersolid transition because 
instead of drop below the possible transition the entropy of 
the solid rather remains constant. Also, the size of the 
anomaly is by four orders of magnitude smaller than one would 
expect for the supersolid transition. The origin of this 
residual entropy remains unclear.

\begin{acknowledgments}
We are grateful to A.\,F.\,Andreev, A.\,V.\,Balatsky,
M.\,Paalanen, R.\,Jochemsen, G.\,R.\,Pickett, A.\,Sebedash,
and G.\,E.\,Volovik for fruitful discussions.  This work was supported by the
EC-funded ULTI project, Transnational Access in Programme FP6
(contract \#RITA-CT-2003-505313) and by the Academy of
Finland (Finnish Centre of Excellence Programmes 2000-2005,
2006-2011, and the Visitors Programme 112401).
\end{acknowledgments}














%































\bibliography{todoshchenko}

\begin{thebibliography}{27}
\expandafter\ifx\csname natexlab\endcsname\relax\def\natexlab#1{#1}\fi
\expandafter\ifx\csname bibnamefont\endcsname\relax
  \def\bibnamefont#1{#1}\fi
\expandafter\ifx\csname bibfnamefont\endcsname\relax
  \def\bibfnamefont#1{#1}\fi
\expandafter\ifx\csname citenamefont\endcsname\relax
  \def\citenamefont#1{#1}\fi
\expandafter\ifx\csname url\endcsname\relax
  \def\url#1{\texttt{#1}}\fi
\expandafter\ifx\csname urlprefix\endcsname\relax\def\urlprefix{URL }\fi
\providecommand{\bibinfo}[2]{#2}
\providecommand{\eprint}[2][]{\url{#2}}

\bibitem[{\citenamefont{Andreev and Lifshitz}(1969)}]{AL}
\bibinfo{author}{\bibfnamefont{A.~F.} \bibnamefont{Andreev}} \bibnamefont{and}
  \bibinfo{author}{\bibfnamefont{I.~M.} \bibnamefont{Lifshitz}},
  \bibinfo{journal}{Zh.\ Eksp.\ Teor.\ Fiz.} \textbf{\bibinfo{volume}{56}},
  \bibinfo{pages}{2057} (\bibinfo{year}{1969}).

\bibitem[{\citenamefont{Meisel}(1992)}]{Meisel}
\bibinfo{author}{\bibfnamefont{M.~W.} \bibnamefont{Meisel}},
  \bibinfo{journal}{Physica} \textbf{\bibinfo{volume}{B 178}},
  \bibinfo{pages}{121} (\bibinfo{year}{1992}).

\bibitem[{\citenamefont{Lengua and Goodkind}(1990)}]{Lengua}
\bibinfo{author}{\bibfnamefont{G.~A.} \bibnamefont{Lengua}} \bibnamefont{and}
  \bibinfo{author}{\bibfnamefont{J.~M.} \bibnamefont{Goodkind}},
  \bibinfo{journal}{J.\ Low Temp.\ Phys.} \textbf{\bibinfo{volume}{79}},
  \bibinfo{pages}{251} (\bibinfo{year}{1990}).

\bibitem[{\citenamefont{Kim and Chan}(2004{\natexlab{a}})}]{Kim}
\bibinfo{author}{\bibfnamefont{E.}~\bibnamefont{Kim}} \bibnamefont{and}
  \bibinfo{author}{\bibfnamefont{M.~H.~W.} \bibnamefont{Chan}},
  \bibinfo{journal}{Science} \textbf{\bibinfo{volume}{305}},
  \bibinfo{pages}{1941} (\bibinfo{year}{2004}{\natexlab{a}}).

\bibitem[{\citenamefont{Kim and Chan}(2004{\natexlab{b}})}]{Nature}
\bibinfo{author}{\bibfnamefont{E.}~\bibnamefont{Kim}} \bibnamefont{and}
  \bibinfo{author}{\bibfnamefont{M.~H.~W.} \bibnamefont{Chan}},
  \bibinfo{journal}{Nature} \textbf{\bibinfo{volume}{427}},
  \bibinfo{pages}{225} (\bibinfo{year}{2004}{\natexlab{b}}).

\bibitem[{\citenamefont{Rittner and Reppy}(2006)}]{Rittner}
\bibinfo{author}{\bibfnamefont{A.~S.~C.} \bibnamefont{Rittner}}
  \bibnamefont{and} \bibinfo{author}{\bibfnamefont{J.~D.} \bibnamefont{Reppy}},
  \bibinfo{journal}{cond-mat/0604528}  (\bibinfo{year}{2006}).

\bibitem[{\citenamefont{Kim and Chan}(2006)}]{Kim06}
\bibinfo{author}{\bibfnamefont{E.}~\bibnamefont{Kim}} \bibnamefont{and}
  \bibinfo{author}{\bibfnamefont{M.~H.~W.} \bibnamefont{Chan}},
  \bibinfo{journal}{cond-mat/0605680}  (\bibinfo{year}{2006}).

\bibitem[{\citenamefont{Dash and Wettlaufer}(2005)}]{Dash}
\bibinfo{author}{\bibfnamefont{J.~G.} \bibnamefont{Dash}} \bibnamefont{and}
  \bibinfo{author}{\bibfnamefont{J.~S.} \bibnamefont{Wettlaufer}},
  \bibinfo{journal}{Phys.\ Rev.\ Lett.} \textbf{\bibinfo{volume}{94}},
  \bibinfo{pages}{235301} (\bibinfo{year}{2005}).

\bibitem[{\citenamefont{Balatsky et~al.}(2006)\citenamefont{Balatsky, Graf,
  Nussinov, and Trugman}}]{Balatsky}
\bibinfo{author}{\bibfnamefont{A.~V.} \bibnamefont{Balatsky}},
  \bibinfo{author}{\bibfnamefont{M.~J.} \bibnamefont{Graf}},
  \bibinfo{author}{\bibfnamefont{Z.}~\bibnamefont{Nussinov}}, \bibnamefont{and}
  \bibinfo{author}{\bibfnamefont{S.~A.} \bibnamefont{Trugman}},
  \bibinfo{journal}{cond-mat/0606203}  (\bibinfo{year}{2006}).

\bibitem[{\citenamefont{Day and Beamish}(2006)}]{Day}
\bibinfo{author}{\bibfnamefont{J.}~\bibnamefont{Day}} \bibnamefont{and}
  \bibinfo{author}{\bibfnamefont{J.}~\bibnamefont{Beamish}},
  \bibinfo{journal}{Phys.\ Rev.\ Lett.} \textbf{\bibinfo{volume}{96}},
  \bibinfo{pages}{105304} (\bibinfo{year}{2006}).

\bibitem[{\citenamefont{Gardner et~al.}(1973)\citenamefont{Gardner, Hoffer, and
  Phillips}}]{Gardner}
\bibinfo{author}{\bibfnamefont{W.~R.} \bibnamefont{Gardner}},
  \bibinfo{author}{\bibfnamefont{J.~K.} \bibnamefont{Hoffer}},
  \bibnamefont{and} \bibinfo{author}{\bibfnamefont{N.~E.}
  \bibnamefont{Phillips}}, \bibinfo{journal}{Phys. Rev.}
  \textbf{\bibinfo{volume}{A 7}}, \bibinfo{pages}{1029} (\bibinfo{year}{1973}).

\bibitem[{\citenamefont{Abraham et~al.}(1970)\citenamefont{Abraham, Eckstein,
  Ketterson, Kuchnir, and Roach}}]{Abraham}
\bibinfo{author}{\bibfnamefont{B.~M.} \bibnamefont{Abraham}},
  \bibinfo{author}{\bibfnamefont{Y.}~\bibnamefont{Eckstein}},
  \bibinfo{author}{\bibfnamefont{J.~B.} \bibnamefont{Ketterson}},
  \bibinfo{author}{\bibfnamefont{M.}~\bibnamefont{Kuchnir}}, \bibnamefont{and}
  \bibinfo{author}{\bibfnamefont{P.~R.} \bibnamefont{Roach}},
  \bibinfo{journal}{Phys.\ Rev.} \textbf{\bibinfo{volume}{A 1}},
  \bibinfo{pages}{250} (\bibinfo{year}{1970}).

\bibitem[{\citenamefont{Grilly}(1973)}]{Grilly}
\bibinfo{author}{\bibfnamefont{E.~R.} \bibnamefont{Grilly}},
  \bibinfo{journal}{J.\ Low Temp.\ Phys.} \textbf{\bibinfo{volume}{11}},
  \bibinfo{pages}{33} (\bibinfo{year}{1973}).

\bibitem[{\citenamefont{Greywall}(1980)}]{Greywall79}
\bibinfo{author}{\bibfnamefont{D.~S.} \bibnamefont{Greywall}},
  \bibinfo{journal}{Phys.\ Rev.} \textbf{\bibinfo{volume}{B 21}},
  \bibinfo{pages}{1329} (\bibinfo{year}{1980}).

\bibitem[{\citenamefont{Landau and Lifshitz}(1980)}]{LL}
\bibinfo{author}{\bibfnamefont{L.~D.} \bibnamefont{Landau}} \bibnamefont{and}
  \bibinfo{author}{\bibfnamefont{E.~M.} \bibnamefont{Lifshitz}},
  \bibinfo{journal}{Statistical Physics, Part I, §62, Pergamon}
  (\bibinfo{year}{1980}).

\bibitem[{\citenamefont{Anderson et~al.}(2005)\citenamefont{Anderson, Brinkman,
  and Huse}}]{Anderson}
\bibinfo{author}{\bibfnamefont{P.~W.} \bibnamefont{Anderson}},
  \bibinfo{author}{\bibfnamefont{W.~F.} \bibnamefont{Brinkman}},
  \bibnamefont{and} \bibinfo{author}{\bibfnamefont{D.~A.} \bibnamefont{Huse}},
  \bibinfo{journal}{Science} \textbf{\bibinfo{volume}{310}},
  \bibinfo{pages}{1164} (\bibinfo{year}{2005}).

\bibitem[{\citenamefont{Castles and Adams}(1975)}]{Castles}
\bibinfo{author}{\bibfnamefont{S.~H.} \bibnamefont{Castles}} \bibnamefont{and}
  \bibinfo{author}{\bibfnamefont{E.~D.} \bibnamefont{Adams}},
  \bibinfo{journal}{J.\ Low Temp. Phys.} \textbf{\bibinfo{volume}{19}},
  \bibinfo{pages}{397} (\bibinfo{year}{1975}).

\bibitem[{\citenamefont{Hanson et~al.}(1976)\citenamefont{Hanson, Seidel, and
  Maris}}]{Hanson}
\bibinfo{author}{\bibfnamefont{H.~N.} \bibnamefont{Hanson}},
  \bibinfo{author}{\bibfnamefont{J.~E. B. G.~M.} \bibnamefont{Seidel}},
  \bibnamefont{and} \bibinfo{author}{\bibfnamefont{H.~J.} \bibnamefont{Maris}},
  \bibinfo{journal}{Phys. Rev.} \textbf{\bibinfo{volume}{B 14}},
  \bibinfo{pages}{1911} (\bibinfo{year}{1976}).

\bibitem[{\citenamefont{van~de Haar et~al.}(1992)\citenamefont{van~de Haar, van
  Woerkens, Meisel, and Frossati}}]{Leiden}
\bibinfo{author}{\bibfnamefont{P.~G.} \bibnamefont{van~de Haar}},
  \bibinfo{author}{\bibfnamefont{C.~M. C.~M.} \bibnamefont{van Woerkens}},
  \bibinfo{author}{\bibfnamefont{M.~W.} \bibnamefont{Meisel}},
  \bibnamefont{and} \bibinfo{author}{\bibfnamefont{G.}~\bibnamefont{Frossati}},
  \bibinfo{journal}{J.\ Low Temp.\ Phys.} \textbf{\bibinfo{volume}{86}},
  \bibinfo{pages}{349} (\bibinfo{year}{1992}).

\bibitem[{\citenamefont{Ruutu et~al.}(1998)\citenamefont{Ruutu, Hakonen,
  Babkin, \mbox{A. Ya. Parshin}, and Tvalashvili}}]{Ruutu}
\bibinfo{author}{\bibfnamefont{J.~P.} \bibnamefont{Ruutu}},
  \bibinfo{author}{\bibfnamefont{P.~J.} \bibnamefont{Hakonen}},
  \bibinfo{author}{\bibfnamefont{A.~V.} \bibnamefont{Babkin}},
  \bibinfo{author}{\bibnamefont{\mbox{A. Ya. Parshin}}}, \bibnamefont{and}
  \bibinfo{author}{\bibfnamefont{G.}~\bibnamefont{Tvalashvili}},
  \bibinfo{journal}{J.\ Low Temp.\ Phys.} \textbf{\bibinfo{volume}{112}},
  \bibinfo{pages}{117} (\bibinfo{year}{1998}).

\bibitem[{\citenamefont{Straty and Adams}(1969)}]{Straty}
\bibinfo{author}{\bibfnamefont{G.~C.} \bibnamefont{Straty}} \bibnamefont{and}
  \bibinfo{author}{\bibfnamefont{E.~D.} \bibnamefont{Adams}},
  \bibinfo{journal}{Rev.\ Sci.\ Instr.} \textbf{\bibinfo{volume}{40}},
  \bibinfo{pages}{1393} (\bibinfo{year}{1969}).

\bibitem[{\citenamefont{Rusby et~al.}(2002)\citenamefont{Rusby, Durieux,
  Reesink, Hudson, Schuster, K$\rm\ddot{u}$hne, Fogle, Soulen, and
  Adams}}]{PLTS2000}
\bibinfo{author}{\bibfnamefont{R.~L.} \bibnamefont{Rusby}},
  \bibinfo{author}{\bibfnamefont{M.}~\bibnamefont{Durieux}},
  \bibinfo{author}{\bibfnamefont{A.~L.} \bibnamefont{Reesink}},
  \bibinfo{author}{\bibfnamefont{R.~P.} \bibnamefont{Hudson}},
  \bibinfo{author}{\bibfnamefont{G.}~\bibnamefont{Schuster}},
  \bibinfo{author}{\bibfnamefont{M.}~\bibnamefont{K$\rm\ddot{u}$hne}},
  \bibinfo{author}{\bibfnamefont{W.~E.} \bibnamefont{Fogle}},
  \bibinfo{author}{\bibfnamefont{R.~J.} \bibnamefont{Soulen}},
  \bibnamefont{and} \bibinfo{author}{\bibfnamefont{E.~D.} \bibnamefont{Adams}},
  \bibinfo{journal}{J.\ Low Temp.\ Phys.} \textbf{\bibinfo{volume}{126}},
  \bibinfo{pages}{633} (\bibinfo{year}{2002}).

\bibitem[{\citenamefont{Keshishev et~al.}(1982)\citenamefont{Keshishev,
  \mbox{A. Ya. Parshin}, and Shal'nikov}}]{KPS}
\bibinfo{author}{\bibfnamefont{K.~O.} \bibnamefont{Keshishev}},
  \bibinfo{author}{\bibnamefont{\mbox{A. Ya. Parshin}}}, \bibnamefont{and}
  \bibinfo{author}{\bibfnamefont{A.~I.} \bibnamefont{Shal'nikov}},
  \bibinfo{journal}{Physics Reviews} \textbf{\bibinfo{volume}{4 A}},
  \bibinfo{pages}{155} (\bibinfo{year}{1982}).

\bibitem[{\citenamefont{Balibar et~al.}(2005)\citenamefont{Balibar, Alles, and
  \mbox{A. Ya. Parshin}}}]{review}
\bibinfo{author}{\bibfnamefont{S.}~\bibnamefont{Balibar}},
  \bibinfo{author}{\bibfnamefont{H.}~\bibnamefont{Alles}}, \bibnamefont{and}
  \bibinfo{author}{\bibnamefont{\mbox{A. Ya. Parshin}}},
  \bibinfo{journal}{Rev.\ Mod.\ Phys.} \textbf{\bibinfo{volume}{77}},
  \bibinfo{pages}{317} (\bibinfo{year}{2005}).

\bibitem[{\citenamefont{Edwards and Balibar}(1989)}]{Balibar}
\bibinfo{author}{\bibfnamefont{D.~O.} \bibnamefont{Edwards}} \bibnamefont{and}
  \bibinfo{author}{\bibfnamefont{S.}~\bibnamefont{Balibar}},
  \bibinfo{journal}{Phys.\ Rev.} \textbf{\bibinfo{volume}{B 39}},
  \bibinfo{pages}{4083} (\bibinfo{year}{1989}).

\bibitem[{\citenamefont{Andreev and \mbox{A. Ya. Parshin}}(1978)}]{AP}
\bibinfo{author}{\bibfnamefont{A.~F.} \bibnamefont{Andreev}} \bibnamefont{and}
  \bibinfo{author}{\bibnamefont{\mbox{A. Ya. Parshin}}},
  \bibinfo{journal}{Sov.\ Phys.\ JETP} \textbf{\bibinfo{volume}{48}},
  \bibinfo{pages}{763} (\bibinfo{year}{1978}).

\bibitem[{\citenamefont{Rolley et~al.}(1995)\citenamefont{Rolley, Balibar,
  Guthmann, and Nozi$\rm\grave{e}$res}}]{Rolley}
\bibinfo{author}{\bibfnamefont{E.}~\bibnamefont{Rolley}},
  \bibinfo{author}{\bibfnamefont{S.}~\bibnamefont{Balibar}},
  \bibinfo{author}{\bibfnamefont{C.}~\bibnamefont{Guthmann}}, \bibnamefont{and}
  \bibinfo{author}{\bibfnamefont{P.}~\bibnamefont{Nozi$\rm\grave{e}$res}},
  \bibinfo{journal}{Physica} \textbf{\bibinfo{volume}{B 210}},
  \bibinfo{pages}{397} (\bibinfo{year}{1995}).

\end{thebibliography}

\end{document}